\documentclass[12pt,letterpaper]{article}
\usepackage{amsmath,amssymb,pgf,pgfarrows,pgfnodes,float,appendix, hyperref}
\usepackage{graphicx}
\usepackage{subfigure}
\usepackage[margin=0.9in]{geometry}

\newcommand{\be}{\begin{equation}}
\newcommand{\ee}{\end{equation}}
\newcommand{\bea}{\begin{eqnarray}}
\newcommand{\eea}{\end{eqnarray}}

\title{{\rm\footnotesize \qquad \qquad \qquad \qquad \qquad \ \qquad \qquad \qquad \ \ \ \ \ \                 RUNHETC-2023-36}\vskip.5in   Holography for Small Values of the Cosmological Constant}
\author{Tom Banks\\
Department of Physics and NHETC\\
Rutgers University, Piscataway, NJ 08854\\
E-mail: \href{mailto:tibanks@ucsc.edu}{tibanks@ucsc.edu}
\\
\\
\\
\\
}
\date{}
\begin{document}
\maketitle

\begin{abstract}  We review recent work on holography for finite area causal diamonds and explore its implications for the description of such diamonds in the AdS/CFT correspondence.  We argue that in empty $AdS_d$ space they're described by a finite dimensional subspace of the space of primary operators at a point, which is not closed under operator product expansions.  They are not related in any simple way to the Type III von Neumann algebras of regions on the boundary or their Type II cross products.  Our description relies on a novel construction of tensor networks, which is manifestly covariant under $SO(d-1)$  and a discrete subgroup of the isometry group of $H^{d-1}$. In a separate section we comment on the Type III von Neumann algebras that appear in models of quantum gravity in $1 + 1$ dimensions. \end{abstract}
\maketitle

\section{Introduction} 

For the past quarter of a century the term "holography" in the context of the quantum theory of gravity has generally referred to the AdS/CFT correspondence.  Most attempts to formulate holographic theories for non-negative cosmological constant (c.c.) have tried to mimic the successes of this framework.  Progress in understanding locality via error correcting codes/tensor networks/Ryu-Takayanagi formulae, and the black hole information problem via "island formulae" and the quantum corrected entropy formula
\begin{equation} S = \frac{A}{4G_N} + S_{matter} , \end{equation} have been assumed to be of quite general applicability.  More recently, following on the work of\cite{LL}, who associated a Type III von Neumann subalgebra of the algebra of CFT operators {\it in the strict $N = \infty$ limit} with a bulk causal diamond,  there have been attempts\cite{Wittenetal} to interpret these formulae in terms of the Type II von Neumann algebra generated from a Type III von Neumann algebra by the cross product construction.  Type II algebras have traces and admit the definition of density matrices and entropy (albeit with additive ambiguities) and so seem suited for this purpose.  It was also proposed to used these constructions for positive cosmological constant and for local regions in more general space-times.   

In fact, there is a considerable amount of evidence that negative cosmological constant is a special case, and that all of the above properties are in fact properties of general quantum field theories (some only in the large N limit), but not of general models of quantum gravity.  Only certain special features are characteristic of QFT models that have "Einstein-Hilbert duals", which is to say that their correlation functions are well approximated by solving the equations of (super) gravity.  What is more, some of the key entropy formulas are NOT satisfied by properly understood black holes for non-negative c.c. .   Furthermore, the results of\cite{LL} are valid only at infinite $N$.  This means that they can be used to all orders in the $1/N$ expansion, whose coefficients are evaluated at infinite $N$, but they are not valid at any finite value of $N$.  They are related to the fact that the construction of bulk operators\cite{BDHMHKLL} gives a local field theory at leading order in the $1/N$ expansion.   All of this should be contrasted with the much older proposal of the Holographic Space-time Formalism (HST)\cite{hst} which also mimicked the rules of Algebraic Quantum Field Theory, but associates a finite dimensional operator algebra with finite area causal diamonds in space-time. 

The key difference between negative and non-negative c.c. is the nature of the quantum state corresponding to the empty maximally symmetric space.   For negative c.c. it is a pure state, which in the AdS/CFT correspondence is the ground state of the CFT.   For positive c.c., we have known since the work of Gibbons and Hawking\cite{gh} that empty dS space is a high entropy mixed state.   We should also have known, from the Schwarzschild de Sitter entropy (SdS) formula, that localized excitations inside dS space {\it reduce} the entropy, and that this is the smoking gun clue to the nature of locality for non-negative c.c..  It is the clue to the relationship between the quantum theory of gravity and bulk quantum field theory, and the explanation of why dS space has a unique temperature.   If one follows it far enough, this clue also leads to the resolution of the firewall paradox\cite{firewall} and the black hole information problem.  

So far, one might think that the case of vanishing c.c. is ambiguous, since it can be approached from both the positive and negative side.   Perturbative approaches to quantum gravity and flat space string theory suggest that it is similar to AdS/CFT: there is a unique Fock ground state.  The non-perturbative framework provided by the BFSS\cite{bfss} matrix models, makes no such statement.   It has no analog of the zero particle state, just a prescription for calculating S matrix elements for supergravitons by taking a large $N$ limit of amplitudes in a theory of $N \times N$ matrices.   Soft supergravitons appear as small matrix blocks with transverse momentum of order $N^{-1/2}$ and there is no control over whether the limiting S matrix is a unitary operator in Fock space.   

When we try to compute the flat space S matrix by directly taking the limit of CFT amplitudes\cite{polchsuss} we in fact find evidence that AdS/CFT leads to a picture of flat space similar to the limit from positive c.c..  Flat space is realized as a small subsystem of the CFT called "the arena".   In modern tensor network language we can think of the arena as the domain of dependence of a single node of the tensor network corresponding to a cut off CFT with an EH dual.   One of the properties of such CFTs is that single nodes of their tensor networks contain a huge number of degrees of freedom.   In a generic finite energy state of the CFT, the single node is entangled with a much larger system, the entire tensor network, so by Page's theorem\cite{page} its reduced density matrix must have very high entropy.  

It's easy to guess what these high entropy states are in the flat space limit (and we've given a hint about it in our remark about the BFSS matrix model): they are states containing infinitely soft (super) gravitons.  We {\it know} that these cause a problem for the Fock space interpretation of S matrix amplitudes, even in perturbation theory, in four dimensions.  Existence of finite Fock space matrix elements in higher dimensions does not guarantee unitarity of the S matrix in Fock space, and both the approach from BFSS matrix models and from AdS/CFT leave the question of the nature of the non-perturbative Hilbert space of Minkowski quantum gravity unanswered.  It is related to correlation functions/S-matrix elements with large numbers of external legs, where that number is correlated with the large $N$ limit that defines the would be Poincare invariant S-matrix.   We also have evidence\cite{raju} that ordinary string perturbation theory breaks down for large numbers of external legs, in any number of dimensions.  In the concluding section we will suggest where the soft supergraviton modifications of the Polchinski-Susskind S-matrix prescription arise in AdS/CFT.

Another reason to expect that flat space has more in common with dS than AdS is the behavior of black holes.   The reference is to the sign of the specific heat, but there is a more illuminating way to illustrate the difference.   Consider a black hole in flat space and a bit of low entropy matter approaching it.  This requires us to think about a causal diamond whose boundary is larger than the black hole horizon.   We would be tempted to write the same entropy formula that we wrote above for AdS space:  the total entropy is the sum of the black hole entropy and the entropy of the matter.   However, we know what happens.  The matter falls into the black hole, forming a new black hole with a {\it slightly} larger radius, and {\it a huge increase in entropy}.  Where did that entropy come from?  If we look at the causal diamond whose boundary is the new black hole horizon, its boundary was, from the bulk QFT point of view, originally in a state corresponding to the QFT vacuum and it's locally still in a state corresponding to the QFT vacuum.  Yet a whole set of q-bits have suddenly become excited.   They must have always been there, but frozen, in the state we described as "bit of matter sitting outside the black hole".   The interaction between the matter and the black hole was mediated by the unfreezing of these q-bits and the process of "falling into the black hole" is the equilibration of the matter degrees of freedom, the black hole degrees of freedom, and the frozen q-bits.  This shows us that the QFT description of the local Hilbert space in the vicinity of a causal diamond boundary is flawed.

Ironically, QFT does detect the presence of a large set of almost degenerate states at a diamond boundary if we view things from an accelerated coordinate system\cite{Unruh}.  This gives rise to an infinite entanglement entropy between the diamond and its exterior\cite{sorkinetal}, which is the essential reason that local operator algebras in QFT are Type III.   This infinity, which lies at the heart of the "firewall paradox"\cite{firewall} was  conjectured in the 1990s\cite{sussugjacob} to be "absorbed into a renormalization of Newton's constant", which implied a radical extension of the Bekenstein-Hawking entropy law from black hole horizons to general causal diamond boundaries.   Note that this suggests both that the nature of the local Hilbert space near a diamond boundary can be properly understood only in a complete theory of quantum gravity, and that there should be no drama at black hole horizons, since both the vacuum and the black hole are viewed as high entropy states.  In QFT, the large number of states detected by an accelerated observer are considered to be very high energy states from the viewpoint of an infalling inertial observer, and only the detailed pattern of entanglement between shortwavelength modes on either side of a null surface prevents that observer from encountering a "firewall".  The picture implied by the black hole entropy formula suggests that there is something fundamentally incorrect about this QFT description.

To have all of this make sense as a consistent whole, we have to think of the Minkowski vacuum as a formal limit of a high entropy state of very large causal diamond, whose holographic screen has area $A_{d-2} R^{d-2}$, where $A_{d-2}$ is one quarter of the area of the unit sphere in $d-2$ dimensions.  In doing so we've chosen a particular Lorentz frame, and an origin of spatial coordinates.  We then define energy, as measured in this frame, in terms of the number of constraints on the holographic Hilbert space.  That is, a state of energy $E$ corresponds to a subspace, such that the combined statistical plus quantum probability of a randomly chosen state being in that subspace is $e^{- 2\pi R E}$.  This definition only makes sense for $ER \gg 1$ and corresponds to states that we will call "localized in the bulk of the diamond".   To make this more precise let us introduce the diamond universe coordinates of\cite{CHM}
\begin{equation}ds^2 = (\cosh \tau + \cosh x)^{-2} (-d\tau^2 + dx^2 + \sinh^2 x d\Omega_{d-2}^2) . \end{equation}  This is one of many inextensible coordinate systems that cover the interior of the diamond, such that one of the time-like lines is the geodesic between the diamond tips and the extremely accelerated lines approach the null generators of its boundary.  On each space-like slice of this coordinate system we can write energy as an integral of contributions from different values of $x$, and the contributions near the boundary at $x =\infty$ are always of order $1/R$.  These are responsible for the bulk of the entropy of the diamond, and according to Cohen Kaplan and Nelson\cite{CKN} (CKN, see below) they are not well understood in QFT.  In brief, CKN argued that the experimental success of QFT was unaffected if one omitted QFT states in the causal diamond defined by the experiment, whose gravitational back reaction would have formed a black hole at least as big as the diamond.  The remaining states have an entropy $\sim (A/4G_N)^{3/4}$ and do not include those contributing to the area law entanglement entropy.

If we give up the field theory model of the near boundary states then neither the arguments for thermal probabilities in dS space from SdS black hole physics (which depend on localization near the geodesic) nor the Gibbons-Hawking calculation, apply to them.  There are of course some states of energy $\sim 1/R$ which are well described by QFT.  These are soft particles with wave functions spread over the entire volume of the diamond.  They make up a negligible fraction of the entropy of the diamond, even though they do not suffer from Boltzmann suppression.  The field theory entropy has a divergence, coming from states of arbitrarily low static energy.  This is nothing but the usual infinite entanglement entropy in QFT of a finite area diamond embedded in a larger space-time, in this case the global dS manifold.  It is state independent and depends on the background geometry only through the area of the diamond's holoscreen.  We will argue that it should be properly interpreted only in a finite theory of QG, and that there it is modeled by a cut-off 2 dimensional CFT living on a strip surrounding the diamond boundary, with a central charge proportional to the area.  The idea that a finite fraction of the entropy is well modeled by bulk QFT has, in the author's opinion, been definitively refuted by the firewall paradox, but old ideas die hard in physics.  States with a finite fraction of the dS entropy are meta-stable SdS black holes.  Thermal QFT states in dS have much less entropy and that entropy is dominated by states of high static energy.

We will now argue that the definition of energy in terms of constraints makes it an asymptotically conserved quantity.  To do so, we have to make a bit of a digression to define what we mean by time evolution.  Here we take a hint from Algebraic Quantum Field Theory\cite{haagbrandeis}, where there are two notions of time evolution, which happen to coincide.  The fundamental notion of time evolution in AQFT is the {\it Heisenberg automorphism} on each local algebra ${\cal A}_{\diamond}$ in the net of Type III von Neumann algebras defining the model\footnote{We specialize to the algebras of causal diamonds, rather than more general regions, a practice followed by some but not all workers in AQFT.}.   When the net of local algebras is realized as subalgebras of the algebra of operators on a Hilbert space with a unique vacuum in a space-time with a global time-like Killing vector, the local Heisenberg automorphisms can be realized by restriction of the action of the global unitary transformation that implements the Killing vector.

In a model of quantum gravity, we retain the notion of local Heisenberg automorphism, but reserve the issue of global implementation for more detailed study.  It depends on the sign of the cosmological constant and is deeply connected with the infrared issues of quantum gravity, which are highly non-trivial for any sign of the c.c..
On the other hand, for finite area causal diamonds, the Covariant Entropy Principle\cite{fsb} tells\footnote{This is the fundamental hypothesis of HST.} us that we can replace Type III algebras by Type $I_N$ algebras where $N \approx e^{\frac{A_{\diamond}}{4G_N}}$.  This implies that Heisenberg automorphisms are unitarily implementable.   {\it It also forces us to face up to a property of Heisenberg automorphisms that was always hidden by the abstract formalism of AQFT.}  They implicitly describe time evolution between time slices like those of the diamond universe coordinates, which stay within the causal diamond.   {\it The Hamiltonians that implement them will be time dependent, even if the global space-time has a time-like Killing vector. }  

\begin{figure}[h]
\begin{center}
\includegraphics[width=01\linewidth]{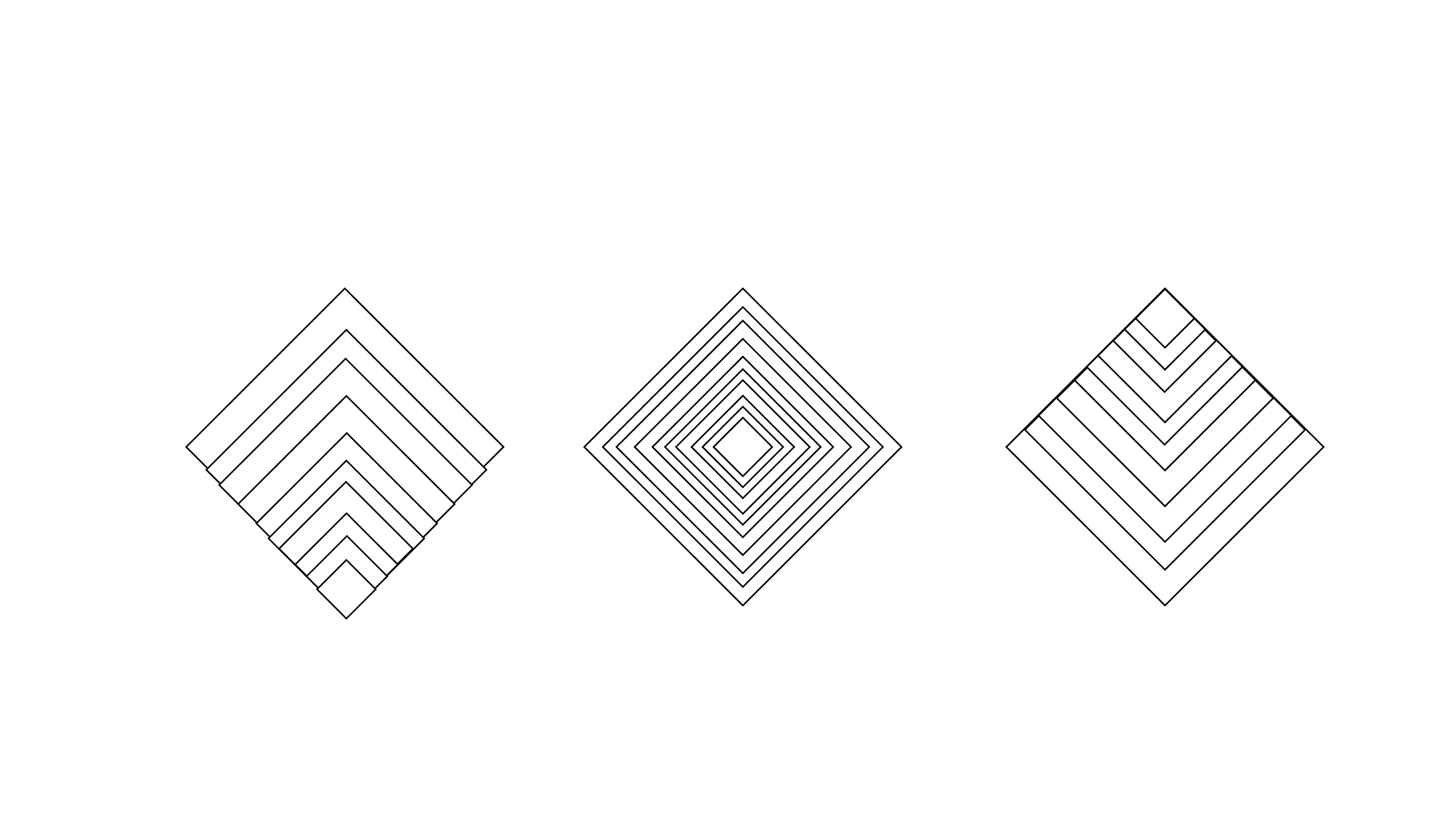}

\caption{Future directed, time symmetric, and past directed nested coverings of a causal diamond.} 
\label{nestedcoversofadiamond}
\end{center}
\end{figure}

Indeed, if we think of a nested sequence of causal diamonds (\ref{nestedcoversofadiamond}) 
corresponding to proper time intervals $[-R, - R + t L_P]$, which cover the large diamond from $[-R,R]$ a Hamiltonian that propagates on time slices interpolating between those diamonds must be time dependent because it corresponds to a sequence of unitary {\it embeddings} of smaller Hilbert spaces into larger ones.  Hamiltonians propagating between the time slices of diamond universe coordinates are better thought of in terms of the time symmetric nesting (\ref{nestedcoversofadiamond}) and must, at each time, split into a sum of a term $H_{in}$ containing only operators inside a given diamond, and another $H_{out}$ describing its exterior.   In QFT we do not see these distinctions because the algebras are all infinite dimensional and do not act on full Hilbert spaces.

The future directed nesting is interesting because of a result in AQFT.  Although Type III VN algebras do not have density matrices, they do have {\it modular operators} which play similar roles.  The modular operator $\Delta$ of the algebra ${\cal A}_{\diamond}$ of a diamond, is constructed with the aid of the QFT vacuum.  It is a positive densely defined operator on the QFT Hilbert space and the one parameter group of unitaries $\Delta^{it}$ maps ${\cal A}_{\diamond}$ onto itself.  Now consider two nested diamonds with infinitesimal proper time separation.   Then $\Delta_-^{-i} \Delta_+^i $ is a unitary that maps the operators localized in the region that is in the larger diamond but not in the smaller one, into themselves.  That is, it implements the Heisenberg time evolution in that nearly null region.

In a CFT there is an even more direct, though less rigorous, argument, relating the modular Hamiltonian of a diamond to Heisenberg time evolution.  Casini, Huerta and Myers\cite{CHM} have argued that one can construct a universal regulator for the leading UV divergence of the entanglement entropy of causal diamond in Minkowski space.  The UV divergence is the volume divergence of the thermal entropy of the CFT on a hyperbolic space, and is manifestly insensitive to perturbations of the CFT Lagrangian by irrelevant operators.  The modular Hamiltonian of the diamond is then proportional to the quantum generator of the conformal Killing vector that leaves the diamond invariant and commutes with $SO(d-1)$, evaluated at the bifurcation surface (the holographic screen) of the diamond.   The diamond universe coordinate $\tau$ parametrizes flow lines of the CKV inside the diamond.   

The proper normalization of the modular Hamiltonian was determined by some remarkable work, more that a decade earlier than that of CHM, by Carlip\cite{carlip} and Solodukhin\cite{solo} (CS).  CS did their work in the context of black holes but the authors of\cite{BZ} pointed out that it applied near the boundary of {\it any} non-extremal causal diamond.   Near any such boundary (past or future), length scales in two dimensions are going to zero, while the transverse dimensions remain large.  By simple scaling, the dominant terms in the Einstein equations become
\begin{equation} R_{ab} = 0 . \end{equation}   These imply that the transverse metric $h_{ij}$ becomes independent of the two nearly null coordinates and that the two dimensional metric
\begin{equation} \tilde{g}_{ab} = g_{ab} + g_{ai}h^{ij}g_{b_j} , \end{equation} is flat.   This remains true in the presence of stress energy that has components only in the nearly null directions, and leads to the famous shock wave equations\cite{tH} for the coordinate transformations that take $\tilde{g}_{ab}$ into $\eta_{ab}$.   Turning to quantum fluctuations, this leads to a canonical derivation of the 't Hooft commutation relations\cite{V29193}, for the coordinate transformations that take the flat metric to Minkowski form.  For CS, the important point was that the linearized fluctuations of the conformal factor of $h_{ij}$ around this classical solution were those of a massless two dimensional scalar field, with a central charge proportional to the area of the transverse metric.   Following the lead of Strominger's\cite{strominger} {\it post facto} "derivation" of the $AdS_3/CFT_2$ correspondence, they took this as a hydrodynamic hint that the correct quantum theory of the near horizon states was a $1 + 1$ dimensional CFT, near a value of the $L_0$ generator determined by the classical solution of the massless scalar field equations.
Applying Cardy's formula for the smoothed density of states, they reproduced the entropy of all black holes, apart from large stable black holes in AdS space (of which more below).

A few remarks are in order here.  
\begin{itemize}
\item The CFTs are chiral.  They live on an interval of longitudinal null coordinate and time in the CFT is light front time on the diamond boundary.  There's one CFT on the past boundary of the diamond and one on the future boundary, related by time evolution in the diamond.
\item  Cardy's formula is a smoothed average of a discrete spectrum.  The real meaning of the CS calculation must be that their classical solution picks out a central point in the spectrum and the entropy formula sums a finite set of states surrounding that point. There's an implicit cut-off on the $L_0$ spectrum.
\item The authors of\cite{BZ} argued that the CS derivation was valid for any diamond boundary (with the exception of extremal black holes and large stable black holes in AdS) and that it implied the universal fluctuation formula $(\Delta K)^2 = \langle K \rangle$ for the modular Hamiltonian.  
\item For Minkowski diamonds, the $L_0$ generator of CS is proportional to the generator of the CKV that CHM claimed was the modular Hamiltonian for CFTs in Minkowski space. So the CS conjecture is the natural conjecture to make for the proper generalization of CHM to models of quantum gravity.   It is consistent with the CEP and agrees with path integral derivations of the entropy of causal diamonds\cite{gh2}\cite{bdf}.  
\item In conformally flat geometries, the CKV generates flows along the time-like diamond universe coordinates.  Together with the general AQFT relation between infinitesimal modular flows and Heisenberg evolution this leads one to conjecture a general formula, which we will discuss below, for the time evolution operator between Planck separated nested causal diamonds, at least for conformally flat geometries.  The actual time coordinates in a large causal diamond that are defined by this notion of evolution are somewhat peculiar.  For a future directed nested covering of the diamond by smaller diamonds, with Planck scale separation between their tips, the time slices are the last Planckian slice of CHM diamond universe coordinates in each diamond in the nest.  The unitary evolution maps are embeddings of smaller Hilbert spaces into larger ones.  There's a similar time reflected picture for past directed nested covers of a diamond.  For the time symmetric nested cover of a diamond it is natural to use the diamond universe coordinates of the largest diamond.  In this case the unitaries operate in the whole Hilbert space but factor into $U_{in} \otimes U_{out}$ on each time slice.
\end{itemize}

In\cite{hilbertbundles} we made a more detailed conjecture about the nature of the CFT of postulated by CS.  It is connected to fluctuations in the unit volume transverse geometry of the holographic screen of the diamond.  Since, in the presence of a cutoff on the discrete spectrum of the $L_0$ generator, it should represent a CFT with finite entropy, it is natural to think of fermionic two dimensional fields.  A. Connes\cite{connes} has shown that $d - 2$ dimensional Riemannian geometry is completely encoded in the Dirac operator on the manifold, so we postulate a spinor field on the holographic screen, built from a finite set of eigenfunctions of the $d-2$ dimensional Dirac operator, each of whose expansion coefficients is a two dimensional fermion field, living on a null strip on the diamond boundary, with an $L_0$ cutoff, centered around a value of $L_0$ determined by the CS calculation.  The values of the two dimensional and transverse cutoffs are determined by the CS calculation.  The two dimensional cutoff is set by the minimal area of the transverse surface for which we believe the semi-classical reasoning of CS.  There doesn't seem to be any other way to determine this cutoff, except perhaps confrontation with experimental data at some time in the distant future.  For that minimal surface we use only zero modes of the transverse Dirac operator.  As we go to larger diamonds, we keep the two dimensional cut-off fixed, and increase the number of transverse eigenmodes to match the area law for entropy.  For $D$ dimensional Minkowski diamonds of large proper time $T$ the number of two dimensional fermion fields scales like $T^{D-2}$.   In dS space, it scales like the diamond area, which remains finite as proper time goes to infinity.  Note that there is a maximal angular momentum state in the Hilbert space, whose angular momentum scales like a power of the area.  The states of very high angular momentum have very low entropy.  

This description of quantum gravity depends on a background space-time and is based on Jacobson's Principle\cite{ted95}: Einstein's equation's are the hydrodynamic equations of the area law.  Given a solution of the hydrodynamic equations, we try to define a quantum system whose hydrodynamics matches it.  That system then provides us with the fluctuation corrections to the classical hydrodynamic equations, as well as more refined quantum observables.  The CS conjecture can be viewed as a refinement of Jacobson's Principle, motivated by the fact that entropy is concentrated on causal diamond boundaries. The hydrodynamics of the conformal factor of the holoscreen metric is that of a two dimensional CFT, and this provides a guess for what the actual quantum theory is.  In\cite{hilbertbundles} we combined this with the HST\cite{hst} conjecture that the variables describing fluctuating transverse geometry were the expansion coefficients of sections of the spinor bundle on the holoscreen into eigenmodes of the Dirac operator, to make a rather specific guess about what CFT was involved.  We argued that it had to be a rather specific form of abelian Thirring model built from the fermion field described above.  This conjecture needs to be fleshed out and investigated more thoroughly.

So far this description has been restricted to the causal diamonds associated with a single time-like trajectory.  We can make it more covariant by specifying a finite dimensional manifold of time-like trajectories, whose diamonds form an open cover of the background manifold.  In many cases the space of time-like geodesics is probably sufficient, but V. Hubeny has pointed out that in AdS space, time-like geodesics do not reach the boundary, so there may be space-times for which one wants a richer space of time-like trajectories \footnote{It should be noted, that we will shortly be describing the rather drastic modifications of this formalism that are necessary to take into account the special case of AdS asymptotics.}.  At any rate, the full model is described as a bundle of quantum systems on a Hilbert bundle over this space of trajectories, with consistency conditions relating the density matrices on overlaps between diamonds.  Details can be found in\cite{hilbertbundles}.  

This general method of analyzing quantum gravity is very explicitly {\it background dependent}.  The background independence of Einstein's equations is viewed as analogous to that of the Navier-Stokes equations, following the seminal paper of Jacobson\cite{ted95}.  Their universality arises because they are the coarse grained expression of the fundamental law relating geometry to quantum information, not because there is a unique underlying model of quantum gravity.  Instead, as explicit examples from string theory have been telling us for decades, there are many mathematically consistent quantum models consistent with this fundamental principle, and we must find the dynamical system that goes with each space-time individually.  There is no such thing as a background independent formulation of quantum gravity.

After this long exposition of the meaning of time evolution in causal diamonds, let us return to the notion of constraints on a large causal diamond as defining an asymptotically conserved energy and the notion of locality in space-times with non-negative c.c..  It is most convenient to think of a time symmetric nesting of diamonds, starting from a large diamond of area $A_{d-2} R^{d-2}$, with an initial number of constrained q-bits of order $ER$.   Evolution is described by a sequence of unitaries of the form $U_{in} \otimes U_{out}$, where, for each interval of proper time, $U_{in}$ acts only on q-bits contained in that interval's causal diamond.   This means that the operator that acts on the constrained variables is only that part of $U_{out}$ that corresponds to large values of the diamond universe coordinates $x$, for the large causal diamond.  But the time scale for evolution of that portion of the system, because of the Milne red shift, is $R$.   Thus, there is not enough time in the causal diamond to change $E$ by more than of order $1/R$.  In the limit $R/L_P \rightarrow \infty$, $E$ is conserved.

In order to insure a degree of locality in space-time physics one must impose locality rules on the holographic screen.  However, we have evidence\cite{lshpss} that the dynamics on the holographic screen of a causal diamond is a fast scrambler of quantum information.  Fischler and the present author have argued\cite{tbwf18} that this is equivalent to requiring that the Hamiltonian be invariant under volume preserving diffeomorphisms of the screen.  The constrained q-bits referred to in the the previous paragraph are interpreted as the vanishing of operator valued measures in annular regions whose volume scales like $ER$\footnote{There are subtleties that need to be worked out.  The measures are constructed from finite sums of Dirac eigensections and so cannot vanish exactly on compact regions.}.  We can now follow these constraints inward to causal diamonds of smaller size.   The tracks of the annular regions of vanishing are the quantum gravity avatars of "particle tracks" in the Feynman diagram description of space-time physics.  In general, the total energy $E$ will consist, initially, of a number of disconnected annuli.  One does not expect to capture all of them in each smaller causal diamond along a particular geodesic.  However, in case that one does, we will eventually encounter a diamond for which
\begin{equation} ER_S \sim R_S^{d-2} , \end{equation} which is to say that the number of constraints is roughly equal to the total number of q-bits.  At this point, the geometrical picture of the constraints on the holographic screen breaks down.   A random state of the q-bits has half turned off and half turned on.   There is no way to distinguish it from a state with half deliberately turned off.   Note also that because of the nature of the redshift in diamond universe coordinates, the time scale for evolution in smaller diamonds is shorter, always about equal to the diamond radius.     We conclude that, in models of this type, if the amount of energy entering into a diamond is less than $\sim R_S^{d-3}$ in Planck units, then the energy will exit the diamond via disjoint annuli, which can be tracked through the future evolution of the system and will leave "particle tracks" through space-time.  On the other hand, once that threshhold is reached, the system becomes a typical equilibrium state of the boundary degrees of freedom.   It is locally indistinguishable from a typical state of an empty diamond, but differs from it because of its history:  it has left a hole of $E(R - R_S)$ constrained q-bits on the holographic screen of the initial diamond, and a corresponding hole on each smaller diamond in the nest.  Those constraints lead to interactions between this newly formed equilibrium, and other localized systems in those larger diamonds\cite{tbwfnewton}, which would not be there for the empty diamond of the same size.

What we have just tried to argue is that our models impose precisely the kind of constraints on the QFT description of particles moving through space-time that were proposed in a prescient paper by Cohen, Kaplan and Nelson\cite{CKN} in 1998.  All extant experiments are done on some near geodesic trajectory in nearly Minkowski space-time over some proper time $T$.  In principle they can explore a causal diamond whose holographic screen has area $4\pi T^2$.  CKN made the observation that the gravitational back reaction of most of the quantum states that QFT attributes to this region, through the expectation value of their stress tensor $\langle T_{mn} \rangle$, would lead to the formation of a black hole larger than or equal to the region.  They imposed a crude UV cutoff to exclude such states, and showed that it had no effect on the precision agreement between QFT and experiment.  The CKN bound was close to being probed by current experiments. Other forms of cutoff\cite{pdetal} put the confrontation with experiment even further into the future.   Imposing the CKN bound, QFT cannot account for the area law entropy of causal diamonds.   The amount of QFT entropy in a diamond necessary to account for all experiments done to date is of order $(A/4G_N)^{3/4}$ - not enough to account for the area law.  The importance of the CKN paper is that it shows that the precise agreement between QFT and experiment is not justification for the statement that QFT in curved space-time is a good approximation to a model of QG.   Our models show how something resembling the Feynman diagram formalism of QFT might emerge from a very different underlying structure, which has the CKN restrictions on QFT built in from the start.

String theorists have felt comfortable taking QFT as a good approximation because of the agreement of string theory amplitudes with small numbers of particles with the general rules of effective field theory.  However, as the CKN argument makes clear, what is really at issue is the breakdown of field theory at {\it high entropy} and there are simply no good arguments that it is valid there, and lots of evidence that quantum field theory is wrong there.   Jacobson has taught us that {\it classical} GR should be a good hydrodynamic description of high entropy states in quantum gravity, but condensed matter physicists know well that the microscopic quantum mechanics of high entropy states of matter is not properly described, even approximately, by quantizing the Navier-Stokes equation.  

In summary then, for non-negative c.c. the generic state of an empty causal diamond of maximal size is highly impure and states with objects localized near the geodesic in the diamond are highly constrained.  Local physics arises from following the pattern of constraints through nested causal diamonds along a rich set of time-like trajectories.

The formalism sketched above can easily be extended to space-times of the form ${\cal M}^d \times K$ where $K$ is compact.  The transverse spinor bundle for Minkowski space is the bundle over $S^{d-2}$ and it is simply replaced by the bundle over $S^{d-2} \times K$.  The really significant change occurs when we take the limit of asymptotically large causal diamonds.  In the appendix we will recall the conjecture\cite{tbir} that the fermionic variables of the HST formalism converge to the generators of the Awada-Gibbons-Shaw\cite{AGS} superalgebra on the null momentum cones Fourier dual to ${\cal I}_{\pm}$, with the constraints defining jets of particles appearing as restrictions on the representations of those algebras.  For massless particles, the AGS generators satisfy  $P_m (\gamma^m )_a^b Q_b (P) = 0$.  For some manifolds $K$ there are stable massive BPS particles.  In this case the limiting algebra will also contain generators satisfying the parity reflected condition $\tilde{P}_m (\gamma^m )_a^b Q_b (P) = 0$, and the mass matrix of BPS particles will appear in the anti-commutation relations of the two types of generators.  

To summarize:  for non-negative small c.c., we argue that the correct formalism is similar to that of Algebraic Quantum Field Theory, with Type I algebras replacing Type III algebras for finite area causal diamonds.  The formalism depends on a choice of background space-time.  The modular Hamiltonian for each diamond is a cut off $1+1$ dimensional CFT with central charge proportional to the area of the diamond's holoscreen, built from fermion fields labelled by the eigenspinors of the Dirac operator on the holoscreen, and a specific form of four fermion interaction\cite{hilbertbundles}.  For space-times conformal to maximally symmetric spaces, the Heisenberg time evolution on a particular set of time slices in nested covers of a diamond is directly related to the modular Hamiltonians.   The state corresponding to "empty maximally symmetric space" has very high entropy and states with bulk localized excitations are defined by constraints that fermion fields vanish on isolated annuli on the holographic screen.

\section{The Case of Negative Cosmological Constant}

As in the case of $\Lambda = 0$ we have to begin by choosing a semi-bounded element of the asymptotic symmetry group and the associated global coordinate system and preferred geodesic on $AdS_d$.   Above we have done this implicitly by working with a particular nested set of causal diamonds. The relevant novel properties of AdS space are that for proper time intervals greater than $\pi R$, causal diamond boundaries hit the boundary of space-time, which is therefore timelike, and have infinite area.   Associated with this is the infinite redshift between the proper time on the central geodesic and time on the boundary.  Any finite global time on the central geodesic is an infinite global time on the boundary.   In the AdS/CFT correspondence we exploit the scale invariance of the asymptotic boundary metric to make the time and space intervals on the boundary finite in AdS radius units.   This means that any finite proper time along the geodesic, or any timelike curve the stays within a finite radius $r$, shrinks to zero in field theory time.   However, in quantum field theory, everything is singular when shrunk to zero time.   It's sufficient to smear fields with smooth functions of time to make their Lorentzian correlators non-singular, but in the absence of time smearing everything is singular.  
This is a point to which we will return: finite proper time local physics in the AdS/CFT correspondence is cutoff dependent, and therefore suffers from a certain amount of ambiguity if one insists on sticking to the continuum CFT.  

The proper way to think about local physics at finite proper time in AdS/CFT is in terms of the Tensor Network/Error Correcting Code formulation (TN/ECC).  For the past decade or so this has been referred to as a "toy model" of AdS/CFT, but in the early literature it was emphasized that physics inside a finite radius in global coordinates should be thought of as taking place in a cut-off version of the CFT\cite{witsuss}.  The TN/ECC formulation is the most elegant formulation yet proposed for the cutoff.  In this paper we will propose some improvements to it, which will incorporate more of the symmetry structure of the theory.

The origins of TN/ECC lie in the real space renormalization group of condensed matter physics, and related work on quantum information theory, and a lot of the properties of tensor networks, like Ryu-Takayanagi formulae and generalized quantum entropy formulae apply to tensor networks for CFTs that do not have EH duals.  Susskind and the present author have emphasized that one of the crucial properties of tensor networks for models with EH duals is that the nodes of the tensor network have Hilbert spaces of exponentially large dimension.  Roughly speaking this is the requirement that the AdS radius be much larger than the Planck scale.   The second requirement is that that the AdS radius be larger than all other microphysical scales, loosely rendered as "AdS radius larger than the string scale", though it's not clear that all models that fail to obey this criterion have a good perturbative string description.  We will incorporate this requirement by explicitly insisting that dynamics inside a tensor network node be described approximately by one of the $M^d \times K^{11-d}$ models of the previous section.  Note that the time dependent proper time Hamiltonians of those models have nothing to do with the AdS Hamiltonian of the boundary field theory.   As we have indicated above, they represent evolution over proper time scales that get squeezed to zero time on the boundary in the field theory continuum limit.  On the other hand, we do expect the values of the conserved energy $E$, defined by constraints, to be related to eigenvalues of the CFT Hamiltonian $K_0 + P_0$.  We'll try to outline this relationship below.

Our first task is to construct the Hilbert space of the tensor network.   The Hilbert space of a single node contains two dimensional spinor fields, transforming in spinor representations of $SO(d-1) \times G_K$ where $G_K$ is the isometry group of the compact manifold $K$.  The maximal angular momentum is a power of the $AdS$ radius in Planck units.  The maximal eigenvalue of the Dirac operator on $K$ is similarly fixed by insisting that the entropy associated with Kaluza-Klein excitations renormalizes the $11$ (or $10$) dimensional Newton's constant to the $d$ dimensional one in the appropriate way.   The anticommutation relations between incoming and outgoing fields on the future boundary of the diamond encode the topology of $K$ via the spectrum of BPS particles. We'll recall these connections in the appendix.

We now construct a configuration of close packed $d - 1$ dimensional balls on the hyperbolic space $H^{d-1}$.  Using the isometry group of the hyperbolic space we can transform the angular momentum centered around the position of each ball at position $\vec{x}_i $ in $H^{d-1}$ into angular momentum w.r.t. the origin of the tensor network.  
We now construct a lattice field theory on each layer of the tensor network, with nearest neighbor couplings, conserving the overall $SO(d-1)\times G_k$ symmetry, plus any discrete symmetries that we are able to.  If we already knew a lattice theory whose critical behavior was described by the CFT on the boundary, we could use the variational methods of Evenbly and Vidal\cite{tnrg} to construct embedding maps, which would systematically take each shell of the network into the next.  According to\cite{tbwfads} these maps should be thought of as time evolution operators evolving between causal diamonds whose holoscreens are the spheres through the centers of two successive shells of the network.  The proper times between successive diamonds are smaller and smaller, reflecting the fact that diamond boundaries reach the infinite boundary of AdS in finite time.  In the limit of infinite shell radius, the embedding map converges to the CFT Hamiltonian on the infinite dimensional CFT Hilbert space.   

A more transparent way to explain the prescription above is to say that the tensor network is a sequence of lattice QFTs, with the lattice points being the centers of successive shells of close packed balls in $H^{d-1}$.  We consider the degrees of freedom at each lattice point to be multiple copies of the cut-off spinor bundle at that point in $H^{d-1}$.  The multiple copies correspond to the eigenfunctions of the Dirac operator on the compact manifold $K$.  We can use the parallel transport in $H^{d-1}$ to construct rotation invariant couplings between spinors at nearest neighbor points on the lattice, constructing, in each shell, lattice models that are exactly invariant under $SO(d-1)$.  

In simple CFTs ,for which corresponding critical lattice models are known, Evenbly and Vidal\cite{tnrg} (EV) have shown that one can construct local entangling maps between nodes in successive shells of the TN, which converge to the CFT Hamiltonian as the shells get infinitely large.  One postulates a TN wave function for the critical lattice Hamiltonian and solves for the best one by using the variational principle.   Fischler and the present author\cite{tbwfads} have interpreted the finite radius entangling maps of EV as proper time evolution along the central geodesic in a particular static coordinate system.  A given shell of the TN is the holographic screen of the causal diamond of a particular proper time interval.  Of course, for realistic CFTs with EH duals, we do not have explicit lattice models whose critical behavior converges to the CFT, but surely such models exist.  

It's clear that the boundary CFT contains an infinite number of copies of the states of a single sub-AdS radius region.  We are now going to argue that one copy is "less than a point" in the CFT.  An orthonormal basis of the states at a point in a CFT is obtained by acting on the conformally invariant vacuum with all primary operators acting at that point.  
States obtained by acting with more general elements of the operator algebra are obtained from these by acting with elements of the conformal group on the primary states.  The Type III subalgebras of regions appear when we smear the latter with smooth functions with compact support in the coset of the stability subgroup of that point.   Most well known examples of AdS/CFT duality with EH duals have moduli spaces which also contain free field points.  For those at least, we know that the list of primary operators includes operators of arbitrarily high angular momentum on the $d-2$ sphere.  It follows from the Covariant Entropy Bound that the space of states in the finite area causal diamond must have a bound on spatial angular momentum.   This is certainly true of our tensor network construction.   We should also note that there is evidence\cite{maldastrom} from continuum AdS/CFT itself that angular momentum in the compact dimensions is bounded.  If the compact and AdS spatial dimensions become indistinguishable below the AdS radius, the same should be true in the AdS spatial directions in a finite diamond.   Thus the operators in a single node of the TN cannot generate the full subspace of primary operators at a point.

Our tensor network construction demonstrates how states of higher angular momentum are built by adding angular momenta of neighboring spheres in the close packed network.  So there must be a "break" in the list of conformal primaries at some conformal dimension that scales like a power of $N$\footnote{The details of where the bound on $SO(d-1)$ angular momentum sets in obviously depend on the volume of the extra dimensions and might depend on more delicate details of the spectrum of the Dirac operator on the compact space $K$.}.  This is a more fine grained requirement on the spectrum of conformal primaries in a CFT with EH dual than the author's previous speculations on this subject\footnote{I first mentioned the conjecture that an EH dual required a gap with only power law growth in the degeneracy of conformal dimensions during Maldacena's lecture on AdS/CFT at Santa Barbara in 1998, and got shouted down.  I marketed this point of view to various people in the field in subsequent years.  It did not become popular until the paper of Polchinski et. al.\cite{heemskerk} put a little more meat on the speculation.}.   The lesson is clear.   {\it The degrees of freedom of a finite causal diamond do not correspond to any well defined sub-algebra of the algebra of field operators of the CFT.}  They might be identified, probably only roughly, with a subset of conformal primaries, but their algebraic properties, and the physics associated with proper times inside the causal diamond is likely to be well defined only in a cut-off, tensor network version of the CFT.   As a consequence of the usual insensitivity of continuum field theory to the details of the cutoff, one is led to wonder how much of that sub AdS radius physics is contained in the continuum CFT at all.  

There are some important clarifications that should be added to this discussion.  Tensor networks/Error correcting codes were originally introduced into AdS/CFT to understand how information deep in the bulk was encoded in a typical eigenstate of the CFT.  The concatenated entanglements of the tensor network ensure that any bulk node is distributed along the boundary in a highly non-local manner, which makes it impervious to large local erasures.   We have been talking about something quite different, namely the relation between the operator algebra of any node of the network, and that of the CFT.  Equivalently, we can think of this as the relationship between states of a node that is nearly at the boundary to states on the boundary.  This relationship is local and we've argued that we need an infinite number of boundary nodes in the cutoff theory, to duplicate the entire primary operator spectrum of the continuum theory.

The second point of possible confusion is with the work of Leutheuser and Liu\cite{LL}, who identify a subalgebra of the $N = \infty$ CFT field algebra with the Type $III_1$ algebra of a bulk QFT in a "finite" causal diamond.  This result is an artifact of the strict $N = \infty$ limit.  Their subalgebra is the algebra of operators in a finite boundary time interval at $N = \infty$.  For any finite $N$ the algebra of operators in a finite time interval is the full algebra of bounded operators on the CFT Hilbert space and is Type $I$ rather than Type III.  It has no commutant.  Correspondingly, the bulk causal diamond to which it corresponds, has infinite area.  This construction is related to the apparent construction of "local" bulk operators\cite{BDHMHKLL} to all orders in the $1/N$ expansion.  

\section{The Errors of Quantum Field Theory in Curved Space-Time}

For all of its history, the subject of quantum gravity has been subjected to the assumption that it was well approximated by a path integral over metrics and matter fields, and that there was a good approximation to that path integral in situations involving long wavelengths and low energies, in which the matter fields were quantized and gravity was treated as a classical background.  There are several good reasons to believe this.  The first of them is the extraordinary success of such an approximation in accounting for experimental data.  The second is the marvelous idea of "effective field theory" (EFT).  Newton's constant clearly defines a fundamental length scale.  Wilson and Weinberg gave us EFT as a tool for analyzing possible corrections to a renormalizable quantum field theory in powers of the ratio between length scales probed by current experiments and those not yet detectable.  Among the attractions of this idea is that explains why everything is described by renormalizable field theory until just before new discoveries are made.  In quantum gravity, the EFT idea seemed to be validated by the Scherk-Schwarz-Yoneya demonstration that string amplitudes had an EFT expansion in terms of an EFT of EH gravity coupled to matter fields.  

Yet from the beginning there were disturbing indications that gravitation did not fit in well with the EFT idea.  Unlike the weak or strong interactions, whose only effect at large distance/time compared to their characteristic scale is in the existence/non-existence of bound structures, gravity dominates the largest scales in the universe.   In the 1970s, more specific problems raised their heads.  I first heard about the cosmological constant problem from Ken Johnson at the Scottish Summer School in 1976.  Hawking posed the problem of unitarity of black hole evaporation in 1974.  In 1998, Cohen Kaplan and Nelson published an extremely important paper\cite{CKN}, already discussed above but worth re-iterating, about quantum gravity and EFT.   Every experiment ever done has been done on a near geodesic trajectory in nearly Minkowski space-time over some period of proper time $T$.   In principle, the device in that experiment can access the states in a causal diamond, to which QFT attributes a Hilbert space of dimension $e^{c V}$, where $V = (T/L_P)^{d-1}$ ($d=4$ for real experiments) if we assume a crude Planck scale UV cut-off.  CKN note that semi-classical estimates indicate that most of these states will have gravitational back reaction that creates a black hole larger than the causal diamond.   They then showed that omitting such states had no effect on the most precise agreement between QFT theory and experiment in $d = 4$.  They did not note this, but the log of the dimension of the Hilbert space of states {\it kept} by CKN, scales like $(A/4G_N)^{\frac{d-1}{d}}$,  where $A$ is the area of the holographic screen of the diamond.   

One of the main aims of the CKN paper was to show that, when the states they omitted were left out of the QFT Hilbert space, the contradiction between the QFT calculation of the c.c. and its actual value was alleviated.  While this is true, the real role of the c.c. had been  revealed earlier in a seminal paper by Jacobson\cite{ted95}: {\it the c.c. should not be thought of as a vacuum energy density.}  Jacobson showed that the double null projection of the Einstein equations on any null vector could be thought of as the hydrodynamic equations of the Covariant Entropy Principle: {\it Every causal diamond in space-time has a von Neumann entropy equal to one quarter of the area of its holographic screen in Planck units.}  The c.c. does not contribute to these equations and so should not be thought of as a hydrodynamic energy density.  This conclusion is reinforced by the AdS/CFT correspondence, which teaches us that the c.c. is a parameter defined by particular models of quantum gravity, which regulates the high energy, high entropy density of states in those models.  The Banks-Fischler conjecture\cite{tbwfdS} about dS space says the same thing for positive c.c..   

The CKN argument, and Jacobson's Principle lead one to reject the notion that QFT in curved space-time is a good approximation to any model of quantum gravity for $d \geq 4$.   The CKN argument says nothing about lower dimension.   What about Jacobson's principle?   The EH action in $1 + 1$ dimensions is purely topological and the simplest dynamical models of geometry require an extra scalar field.  Considerations of dimensional reduction on a fixed transverse profile from higher dimensions suggests that the coefficient $S$ in the term $S \sqrt{-g}R$ be thought of as an entropy field.   An analog of Jacobson's argument\cite{bdz} derives the equations of a general "dilaton gravity coupled to matter" as the hydrodynamic equations of this system.   Now however, since there is no independent geometrical measure of $S$, nothing like the CKN bound emerges.  One merely finds that $S$ is an auxiliary field.  It has a background profile, which defines the asymptotics of the space-time in which the quantum dynamics takes place, and a quantum piece that is completely determined by matter fields.  The nature of the matter system is determined by {\it a priori} entropy constraints on the system.   

The most well understood of these systems is the exactly soluble Type 0B string\cite{emiletal}\footnote{There are a number of similar related models to which less attention has been devoted.}, which is dual to both a large N bosonic matrix model and a non-relativistic fermionic field theory\cite{matrixmodelandmoore}.  This model has no black holes, and an infinite number of conservation laws, but belongs under the rubric of quantum gravity because it is a string theory to all orders in perturbation theory.   The Hilbert space is infinite dimensional, as befits a scattering theory, and an elegant transformation\cite{AKK} converts the model into a scattering theory of two types of relativistic Weyl fermions with a scattering matrix whose core is given by the 't Hooft commutation relations for eikonal gravitational scattering!   The asymptotic Hilbert space is that of a QFT and it has Type III subalgebras, which appear to be perfectly gauge invariant.   The logarithmically divergent boundary entropy of a finite causal diamond in the relativistic presentation of the fermion Hilbert space is the usual sort of "universal" log one encounters in renormalizable field theory and one does not even need the CHM trick to justify taking it seriously.   I am not sure what the Type II cross product entropies for the Type III subalgebras in this model mean, or what their relevance is to the rather straightforward physics.

A somewhat less well known and perhaps more controversial variation on the OB string models was explored in\cite{lindil12}.  One combines $N$ copies of the OB model with $N \gg 1$ and adds interactions, which in the non-relativistic fermion language, are concentrated near the top of the inverted harmonic oscillator potential.   These interactions break almost all the conservation laws of the model and create meta-stable excitations with many of the properties of linear dilaton black holes.   In particular, they have high entropy, which can be roughly estimated from the Hartree approximation to the dynamics.  The construction of these models uses the local field algebra of the original model in an essential way.  None of it is viewed as a gauge artifact.   The models describe something that looks a lot like black hole formation and evaporation, if one just looks at configurations of the graviton and dilaton fields, in a manifestly unitary manner. The semi-classical phenomenology of the "black-holes" in these models is similar to the "thunderpop" conjecture of\cite{rst}.

The infinite entropy of these systems can be understood as an extrapolation of finite entropy linear dilaton black holes in four dimensional compactifications of string theory\cite{strometal}.   In those configurations the string coupling asymptotes to a finite value and the linear dilaton throat attaches on to an asymptotically flat region.   The CGHS models\cite{CGHS} push the asymptotic region off to infinity and the entropy in the throat becomes infinite at the same time.  From the point of view of four dimensional Einstein frame this is because the area of the transverse sphere at the mouth of the throat is allowed to become infinite.  Type 0B string theory can be viewed as a soluble version of these theories which does not really have an EH dual.    The difference between the versions of these models that have finite entropy black hole excitations and those which do not, does not show up in any finite order of the string perturbation expansion.   It does not seem plausible that leading order entropy calculations based on the idea that the gravitational action is the semi-classical approximation to the full dynamics, could properly distinguish between these models.   If instead we think of gravity as describing the hydrodynamics of the models, then it calculates black hole entropy correctly for the models that have black holes, and simply says nothing about those that don't.   The gravitational action by itself cannot, in these cases at least, tell us what the real physics of the system is.  For the 0B string, which is in the extreme quantum regime, the gravitational action tells us almost nothing about the true dynamics and its black hole solutions are irrelevant.  For the models of\cite{lindil12} it gets the coarse grained hydrodynamics right but misses all the details of the quantum amplitudes and the differences between instantiations of the SYK dynamics.

A more controversial example is the resolution of the degeneracy of extreme non-supersymmetric 4 dimensional Reissner-Nordstrom black holes.  These are finite entropy systems.  They are modeled, semi-classically, by JT-gravity\cite{JT} or JT-gravity coupled to conformal $1+1$ dimensional field theory.   It is widely believed that the spectrum of a typical extremal RN black hole is similar to that of a typical member of the SYK\cite{SYK} ensemble of finite entropy quantum systems.   However, the treatment of the SYK models
by the usual algorithms of the AdS/CFT correspondence do not agree with the semi-classical picture of the physics.  Instead they seem to show a model with an AdS radius of microscopic size:   there is no large gap in dimension between the excitation described by the JT gravity Lagrangian and the rest of the bulk fields.  Two sets of recent papers have attempted to find a quantum gravity description more in tune with the semi-classical physics.  The first approach\cite{TypeII} uses an abstract algebraic description which is only supposed to be valid to leading order in an expansion in Newton's constant.  It can calculate entropies up to a universal additive constant, but does not say much about particular bulk correlation functions and local time evolution.   The second approach\cite{bdz} uses the AKK\cite{AKK} map of relativistic fermions into a large N matrix model, to provide a detailed finite entropy density matrix ansatz, which reproduces several aspects of the bulk physics of fermions propagating in the bulk RN throat.  In particular, it captures the relationship between the JT scalar, which measures entropy in causal diamonds in the throat, and the infall time of fermions to the horizon.  The model also incorporates the SYK interactions explicitly, so that it will settle down to an equilibrium state consistent with SYK spectral statistics.   But it captures the transient bulk dynamics of massless fields escaping from the horizon temporarily and being reflected from the AdS boundary.  It is, one hopes, the first step towards a more complete model which would incorporate coupling to the asymptotically flat region and the study of decay of near extremal black holes back to the extremal limit.  It is not clear to the present author how the methods of\cite{TypeII} could be extended to capture this transient semi-classical physics.

For non-negative c.c. there is a version of the CKN argument in three dimensional space as well.  Einstein's equations coupled to classical particle trajectories are exactly soluble and only allow a finite number of particles, with bounds on what one would have called their Mandelstam invariants (in flat space).  One again comes to the conclusion that QFT coupled to classical gravity is not a good approximation to any sort of quantum gravitational model with these boundary conditions.  A proposal for a quantum theory of three dimensional dS space, consistent with these classical solutions and with several general principles, appeared in\cite{satbpdwf}.    For negative c.c. a similar analysis leads one to BTZ black holes and the $AdS_3/CFT_2$ correspondence.   A contradiction with the zero c.c. limit is avoided (empirically) by the fact that all well understood examples of $AdS_3/CFT_2$ with EH duals, also have at least two compact dimensions with size of order the AdS radius.   The limiting flat space theory has five or more Minkowski dimensions\footnote{Conversations with E. Martinec and S. Sethi have brought up the possibility of examples where the large compact dimensions do not imply extra flat dimensions in the limit.  An analog would be extreme Reissner Nordstrom black holes of large charge.  These appear to have four large dimensions, but since no angular momentum is allowed on the sphere, they do not approach $4$ dimensional Minkowski space, but remain a system with one effective spatial dimension.}.  

\section{Conclusions}

Holography has a very different character for negative and non-negative c.c., stemming from the entirely different nature of the quantum state corresponding to the maximally symmetric space-time.   For non-negative c.c. it is a high entropy state, with entropy going to infinity in the zero c.c. limit.  This conclusion follows when zero c.c. is approached from either positive or negative directions.   The nature of the zero c.c. Hilbert space, and whether there is a unitary scattering operator on it, {\it in any number of dimensions}, remains to be resolved.   What is clear is that for any non-negative value of the c.c., localized excitations inside a causal diamond are constrained states of the holographic degrees of freedom on the boundary.   The precise way in which this works was first suggested in\cite{bfm} and has been elaborated over the years\cite{hst}.  The best current understanding is summarized in\cite{hilbertbundles}.   

For negative c.c. the maximally symmetric space is a pure quantum state, the CFT vacuum, and localized excitations in the bulk are created by acting on the vacuum with local fields on the boundary.   The general consensus has been that bulk locality is somehow captured by the tensor network/error correcting code construction.   There are three related problems with this idea.  The first is that TN/ECC works for an arbitrary CFT, which does not have to have a large number of degrees of freedom and certainly not an EH dual.  The second is that all known AdS/CFT dual pairs with EH duals have large compact dimensions, with radii comparable to the AdS radius.  TN/ECC does not give us a local construction of the compact dimensions, nor show the approximate rotation symmetry relating them to the AdS spatial directions.   Finally, and obviously related to the other two, TN/ECC does not provide a description of locality on scales smaller than the AdS radius.

For a long time, both Susskind and the present author have argued that an essential part of any model with an EH dual is that the nodes of the tensor network must have very large Hilbert spaces.   In this paper, we've made this more quantitative.   The log of the dimension of the node Hilbert space should scale like $N^{D-2}$ where $D$ is the total number of AdS plus large compact dimensions and $N$ is the AdS radius in Planck units. We've also remarked that there should be an upper bound on the angular momentum on the $d-2$ sphere carried by the states in this Hilbert space, which scales like a power of $N$.   The CFT Hilbert space is a tensor product of an infinite number of copies of the node Hilbert space.  It is generated by acting with conformal primary operators on the CFT vacuum and then acting with conformal generators on those primary states.  In all known examples of models with EH duals, there are conformal primaries with arbitrarily large angular momentum on the $d-2$ sphere.  Thus, while it is clear that we should think of the Hilbert space of a single tensor network node near the boundary of AdS as localized at a point on the CFT sphere, it cannot make up the full space of conformal primaries.   We've shown how to glue together the $SO(d-1)$ angular momenta of individual nodes to make lattice models that are actually $SO(d-1)$ invariant.  The full space of conformal primaries at a point is gotten by tensoring together an infinite number of copies of the Hilbert space of a single node.  Thus the information about sub-AdS scale local physics is very well hidden in the continuum CFT.   Type III sub-algebras of the CFT operator algebra and their Type II cross products are too coarse grained to have access to it.  Indeed, it's not entirely clear that the local physics is well defined without the UV cutoff implicit in a tensor network.

The Susskind-Polchinski\cite{polchsuss} prescription does seem to allow one to compute certain Fock space matrix elements of the Minkowski S-matrix as limits of CFT correlators, though it has never been extended to states with fixed momenta in the dimensions transverse to AdS space.  It also cannot deal with infinite numbers of soft massless particles\cite{tbwfads2} and was based on the idea that the Minkowski Hilbert space was Fock space.  This is contradicted by viewing it as the limit of either sign of non-zero c.c. .   In the AdS context, a possible origin of the "infinite soft graviton" part of the Minkowski Hilbert space could be connected correlators of the form
\begin{equation} \langle O_{PS} (1) O_{PS} (2) O_{PS} (3) O_{PS} (4)  O_1 (z_1) \ldots O_n (z_n) \rangle , \end{equation} with arbitrarily large $n$.   Here the first four operators create lines in Witten diagrams that are focussed on "the arena", while the others are random operators of arbitrary dimension, whose contribution to Witten diagrams is diffusely spread over the bulk of AdS space.   There will certainly be diagrams where the two kinds of operators are connected by exchange of soft particle lines in the limit as the AdS radius goes to infinity.   Unitarity sums in the CFT Hilbert space of course include all states created by finite numbers of fields acting on the vacuum, and it might be that the states necessary to make the Fock space S matrix unitary are limits of linear combinations of states like these.   

There have been recent speculations\cite{Waldetal} that there is no unitary S matrix for gravity in four dimensions.  In higher dimensions, the cases treatable by BFSS matrix theory make this seem implausible.   There {\it is} a unitary S matrix for finite $N$ and one just has to prove that a limit exists.  Similar, though perhaps somewhat weaker, remarks would appear to apply to the construction of the scattering matrix from AdS correlators.  There are no examples of either BFSS constructions or reliable AdS/CFT limits that lead to exactly four flat dimensions.   Of course, for practical purposes, all one ever measures are inclusive cross sections, which are finite.

One of the most important conclusions of this analysis is that bulk QFT coupled to gravity is a good approximation to real models of quantum gravity only in certain $1 + 1$ dimensional examples.  It is likely, though the details of this remain to be worked out, that this can be viewed as a consequence of the general Carlip/Solodukhin principle, which states that the dynamics on any causal diamond boundary is that of a cut-off $1 + 1$ dimensional CFT.   In a typical space-time the rules of quantum gravity that we have outlined tell us that in addition to solving these boundary CFTs, one must follow the evolution of the transverse geometry of the diamond boundary to see how the CFT itself evolves, but these $1 + 1$ dimensional models describe throats of extremal black holes, where the transverse geometry is frozen.  Thus, the description of $1 + 1$ dimensional QG models in terms of QFT in curved space-time {\it may} in the end be derived from the general Jacobson-Carlip-Solodukhin picture presented here.

Two final remarks about the view of quantum gravity presented here are necessary.  The first is that it is very much {\it background dependent}.  The idea of background independence in QG stems from the universality of Einstein's equations, which from the present viewpoint is analogous to the universality of the Navier-Stokes equations, and says nothing about the universal nature of the underlying quantum physics, but is a result of the universal connection between space-time geometry and quantum information.   The second point is specific to quantum physics in asymptotically dS space-time - a system with a finite number of quantum states.   Quantum measurement theory requires an experimenter to have a separate semi-classical collective variable, whose quantum fluctuations are small, and whose quantum histories decohere, for each q-bit of the quantum system she/he wants to measure.   The only quantum systems theorists know how to construct with such rich sets of collective variables are QFTs.  The CKN bound shows us that the number of QFT q-bits in dS space is much smaller than the total number of q-bits in the system, and the number of collective variables is obviously still smaller.  As a consequence, there is a theoretical upper limit to the precision with which any quantum theory of dS space could ever be checked by experiment.    

In addition to this, measurement theory in dS space has the following paradoxical feature.  A more complex measuring apparatus, with more collective variables, will undoubtedly have larger mass.  This implies that it is defined in a more constrained subspace of the Hilbert space of the system, so cannot possibly measure all of the mathematical quantum observables in one's theoretical model.  The harder one tries to capture the quantum essence of dS space, the more hidden it becomes.

All of these observations make the construction of elaborate mathematical theories that involve limiting processes and infinities an exercise in futility.  We should concentrate our efforts on finding those aspects of the theory that can make predictions about experiments and observations that can actually be done.  The clue to doing this was actually proposed long ago\cite{cosmoobs}.  In the limit of vanishing c.c. most of the states in the dS Hilbert space become states of zero momentum supergravitons in a model of QG in Minkowski space.  We know that inclusive cross sections with an energy cutoff on energy emitted into supergravitons are finite in the Minkowski limit. The paper\cite{cosmoobs} suggested that there were finite time transition probabilities in dS space, which would converge to those finite inclusive cross sections in the zero c.c. limit, and that these would be the proper observables for a model of dS quantum gravity.  The mathematical definition of these quantities will undoubtedly lack precision, but it only needs to be as precise as the {\it a priori} limits on the precision of measurements in dS space that we have just discussed.

\begin{center}
Acknowledgements
\end{center}

The work on this paper was begun while the author was attending the Benasque Workshop on Quantum Gravity: New Perspectives from Strings and Higher Dimensions. The author would like to thank the organizers, R. Emparan, V. Hubeny, and M. Rangamani, and the staff of the Centro de Ciencias Pedro Pascual for their hospitality during his stay there.  He would also like to thank E. Martinec, S. Sethi, and E. Rabinovici for comments on an earlier draft, which led to major improvements in this work. The work was supported in part by the U.S. Department of Energy under grant DE-SC0010008.

\vfill\eject
\begin{center}
{\bf Appendix}
\end{center}
The variables describing a finite causal diamond in Minkowski space are sections of the spinor bundle $q_m ({\bf \Omega})$ on $S^{d-2}$, with an angular momentum cutoff reflecting the size of the holographic screen.  If the space-time has an extra compact factor space, they will carry an extra label denoting eigensections of the spinor bundle on that manifold.  Each of these variables is a cut-off massless one dimensional fermion field.   States localized in the bulk of the diamond are constrained by the vanishing (in some fuzzy sense that is not yet completely understood) of $q_m$ on annular regions with area that scales like the $\frac{1}{d-2}$ power of the volume of the sphere.  The coefficient of that scaling law is, in the limit of large volume, the asymptotic energy of a localized jet of particles concentrated inside that annulus.  

The asymptotics are defined by taking the radius to infinity.   In this limit, the scaling law of the constrained area is like that of the null momentum on the cone which is the image of the sphere under the conformal group.  Thus its clear that the system wants to be defined in such a way that it has a Lorentz invariant limit, with the energy being the time component of a $d$ vector.  The transversality of the spinor variables is equivalent to the constraint
\begin{equation} P_m (\gamma^m )_a^b Q_b^I (P) = 0 , \end{equation} on Lorentz covariant spinors on the null cone.  $I$ labels the basis of eigensections of the compact Dirac operator.  

Operators of just this type were derived from SUGRA by Awada, Gibbons and Shaw\cite{AGS}.  They satisfy the algebra
\begin{equation} [\bar{Q}_a^I (P), Q_b^J (P^{\prime})]_+ = \delta (P\cdot P^{\prime}) \delta^{IJ}  (\gamma^m)_{ab} P_m . \end{equation}
Up to overall normalization, this is the same algebra that follows from our formalism, for the zero modes of the two dimensional quantum fields.  We will discuss the rest of the two dimensional field theory below.  

Some compact manifolds support massive BPS particles, which arise in string/M theory from branes wrapped around supersymmetric cycles.  These can be described by adding a second set of asymptotic generators.   If $P_m = (E,\vec{P})$, define $\tilde{P}_m = (E, \vec{-P})$, and generators $\tilde{Q}_a^I$, such that 
\begin{equation} \tilde{P}_m (\gamma^m)_a^b \tilde{Q}_b^I (P) = 0 . \end{equation}   Then the mass matrix of BPS particles will appear in the anti-commutation relations between $Q_a^I$ and $\tilde{Q}_a^I$.   There may also be stable massive BPS-anti-BPS bound states classified by discrete K theory charges (or their M theory analogs).   These do not appear in the algebra but show up dynamically in the scattering matrix of the BPS particles.  

For finite causal diamonds, the role of the two dimensional field theory is to describe quantum fluctuations of the transverse geometry of the holographic screen.  The time scale for these dynamical fluctuations is $1/R$ the radius of the screen, and they become frozen as the screen approaches infinity.  The hypothesis we have made in this appendix is that those degrees of freedom decouple from the contrained degrees of freedom describing particle jets.  We have simply omitted them from the asymptotic Hilbert space.   In\cite{hilbertbundles} we made a detailed conjecture about the dynamics of the two dimensional CFT on the holoscreen: it was an abelian Thirring model.   Perhaps that conjecture can eventually be used to prove the decoupling postulated here.

\end{document}